\documentclass[notitlepage,twocolumn,letterpaper,natbib,aps,prd,amsmath,amsfonts,nofootinbib,preprintnumbers,superscriptaddress,secnumarabic,groupedaddress]{revtex4-2}
\pdfoutput=1
\usepackage{amssymb,amsmath,latexsym,mathrsfs}
\usepackage{url}
\usepackage{enumitem}
\usepackage{graphicx}
\usepackage{booktabs}
\usepackage[usenames,dvipsnames]{color}
\usepackage[breaklinks,colorlinks,urlcolor=blue,citecolor=blue,linkcolor=magenta]{hyperref}
\usepackage{multirow}
\usepackage{float}
\usepackage{cases}
\usepackage{blindtext}
\usepackage{pifont}
\usepackage{hhline}

\newcommand{\secskip}{\vspace{6pt}}

\usepackage{xcolor}
\definecolor{linkcolor}{rgb}{0.0, 0.47, 0.75}
\definecolor{citecolor}{rgb}{1.0, 0.5, 0.0}

\hypersetup{
  linkcolor  = linkcolor,
  citecolor  = linkcolor,
  urlcolor   = linkcolor,
  colorlinks = true
}

\begin{document}

\title{Constraints on Global Symmetry Breaking in Quantum Gravity\\ from Cosmic Birefringence Measurements}

\author{James Alvey}
\email{james.alvey@kcl.ac.uk}
\thanks{ORCID: \href{https://orcid.org/0000-0003-2020-0803}{0000-0003-2020-0803}}
\affiliation{Department of Physics, King's College London, Strand, London WC2R 2LS, UK}

\author{Miguel Escudero Abenza}
\email{miguel.escudero@tum.de}
\thanks{ORCID: \href{https://orcid.org/0000-0002-4487-8742}{0000-0002-4487-8742}\vspace{6pt}}
\affiliation{Physik-Department, Technische Universit{\"{a}}t, M{\"{u}}nchen, James-Franck-Stra{\ss}e, 85748 Garching, Germany}

\preprint{KCL-PH-TH 2021/36 $\,\,\,$ TUM-HEP 1342/21}

\begin{abstract}
\noindent All global symmetries are expected to be explicitly broken by quantum gravitational effects, and yet may play an important role in Particle Physics and Cosmology. As such, any evidence for a well-preserved global symmetry would give insight into an important feature of gravity. We argue that a recently reported $2.4\sigma$ detection of cosmic birefringence in the Cosmic Microwave Background could be the first observational indication of a well-preserved (although spontaneously broken) global symmetry in nature. A compelling solution to explain this measurement is a very light pseudoscalar field that interacts with electromagnetism. In order for gravitational effects not to lead to large corrections to the mass of this scalar field, we show that the breaking of global symmetries by gravity should be bounded above. Finally, we highlight that any bound of this type would have clear implications for the construction of theories of quantum gravity, as well as for many particle physics scenarios.
\end{abstract}

\maketitle

\emph{Introduction.---} Global symmetries are ubiquitous in Particle Physics. Sometimes these symmetries appear accidentally as a consequence of gauge symmetries and particle content -- such as baryon and lepton number in the Standard Model -- but they have also been used to address certain open phenomenological issues. For example, they have been invoked in the Peccei-Quinn solution to the Strong CP problem~\cite{Peccei:1977hh,Peccei:1977ur}, or to understand the masses and mixing pattern of fermions~\cite{Babu:2002dz}. In the context of Cosmology, global symmetries can also provide a mechanism to stabilise the dark matter of the Universe~\cite{Pagels:1981ke}, or to explain the ultra-light nature of a scalar field playing the role of dark energy~\cite{Frieman:1995pm}.

On the other hand, it is widely believed that all global symmetries are explicitly broken by gravitational effects~\cite{Kallosh:1995hi,Witten:2017hdv}\footnote{Indeed, the non-existence of exact global symmetries in quantum gravity is one of the key aspects of the swampland program~\cite{Vafa:2005ui}, see e.g.~\cite{Palti:2019pca,Brennan:2017rbf}.}. These general expectations are supported by theoretical calculations illustrating the explicit breaking of global symmetries by sources such as black holes~\cite{Banks:2010zn} or wormholes~\cite{Giddings:1987cg}. In addition, it has been shown that certain concrete theories of quantum gravity do not admit exact global symmetries, including holography~\cite{Harlow:2018jwu} and string theory~\cite{Polchinski:1998rr}. 

Despite this recent theoretical progress, there are no definitive conclusions regarding the explicit breaking of global symmetries by gravity. In particular, as of yet, it is unclear what the actual sources, mechanisms or magnitude of this breaking might be in reality\footnote{For example, in contrast to the expectations from string theory, in alternative scenarios such as asymptotically safe quantum gravity, there are some indications that global symmetries  might be preserved~\cite{Eichhorn:2020sbo}.}. In this letter, we focus on the last aspect, and use a cosmological measurement to motivate how one could set an \emph{observational} upper limit on the breaking of global symmetries by gravitational effects. Such a bound would have clear implications both on the construction of theories of quantum gravity, as well as particle physics scenarios that rely on global symmetries.

More specifically, we argue that a measurement of cosmic birefringence could be used to constrain the magnitude of global symmetry breaking. This is particularly motivated by a recently reported detection of cosmic birefringence in Planck legacy data~\cite{Minami:2020odp}, and by the original description of this phenomenon in terms of pseudoscalar fields in cosmology~\cite{Carroll:1989vb,Harari:1992ea,Carroll:1998zi,Lue:1998mq,Liu:2006uh}. These pseudoscalar fields are naturally very light as a result of an approximate global symmetry~\cite{Frieman:1995pm}, and are also a good fit to the birefringence measurement~\cite{Fujita:2020aqt,Fujita:2020ecn,Takahashi:2020tqv,Jain:2021shf}. This second feature arises as a result of a parity violating coupling to electromagnetism, something that is not possible for scalar fields. Here, we point out that this measurement could represent the first indication of a well-preserved (although spontaneously broken) global symmetry in nature. In addition, we show how to use it to set an observational bound on gravitational global symmetry breaking.

\secskip
\emph{Cosmic Birefringence.---} Planck data has provided us with a deep understanding of the physics of the Cosmic Microwave Background (CMB)~\cite{Aghanim:2019ame}. Importantly, it has been used to obtain detailed insight into the dynamics and content of the Universe, as well as a number of other phenomena~\cite{Aghanim:2018eyx}. In particular, CMB data can be used to constrain parity violation on cosmological scales -- cosmic birefringence being one such example which involves photons. Physically, cosmic birefringence corresponds to a rotation of the linear polarisation of electromagnetic waves as they propagate along a particular line of sight from their source. In observational terms, the existence of this effect directly translates into an effective power in the EB polarisation spectrum of the CMB, which can be parametrised by the cosmic birefringence angle $\beta$. The detection of a non-zero value for $\beta$ would then correspond to an observation of parity violating interactions in cosmology. Very recently, Minami and Komatsu~\cite{Minami:2020odp} have analysed Planck polarisation data and report a measurement of this angle $\beta = 0.35^{\circ} \pm 0.14^{\circ}$, which excludes zero with a significance of $2.4\sigma$.

Perhaps the best motivated possibility to explain the birefringence measurement is the existence of a dynamical pseudoscalar field $\phi$ which interacts with electromagnetism via a coupling of the form $(1/4)g_{\phi\gamma\gamma}\phi F_{\mu\nu}\tilde{F}^{\mu\nu}$~\cite{Carroll:1989vb,Harari:1992ea,Carroll:1998zi,Lue:1998mq}\footnote{We note that there is plenty of theoretical motivation for considering such light scalar fields interacting with electromagnetism. For example, a variety of pseudo-Goldstone bosons coupled to the SM gauge fields are expected to arise from string compactifications~\cite{Svrcek:2006yi,Arvanitaki:2009fg}.}. In this scenario, the temporally-varying scalar field $\phi$ permeates the Universe, acting as a parity violating medium for CMB photons. In particular, the coupling of $\phi$ to electromagnetism results in different dispersion relations for the left and right circular polarisations, which in turn rotates the linear polarisation plane~\cite{Carroll:1989vb,Harari:1992ea,Carroll:1998zi,Lue:1998mq,Liu:2006uh}. Explicitly, the cosmic birefringence angle $\beta$ can be computed as,
\begin{equation}\label{eq:beta_eq}
   \beta = \frac{1}{2}g_{\phi \gamma \gamma}\left(\phi_0 - \phi_\star\right),
\end{equation}

\noindent where $\phi_0$ and $\phi_\star$ are the values of the scalar field today and at the time of recombination respectively. Given a coupling $g_{\phi\gamma\gamma}$, and once the dynamics of the scalar field have been calculated, this value of $\beta$ can be compared to the measurement given above. In practice, the dynamics are controlled by the interplay between the Hubble rate $H$ and the potential of the scalar field $V(\phi)$, as dictated by the Klein-Gordon equation,
\vspace{0.1cm}
\begin{equation}\label{eq:KG_eq}
   \frac{\mathrm{d}^2\phi}{\mathrm{d}t^2} + 3H(t) \frac{\mathrm{d} \phi}{\mathrm{d}t} + \frac{\mathrm{d}V(\phi)}{\mathrm{d}\phi} = 0\,.
\end{equation}
\vspace{0.05cm}

\noindent Within the context of cosmic birefringence, two key scales relevant to the scalar field dynamics are the Hubble rate today $H_0 \sim 10^{-33} \, \mathrm{eV}$, and its value at recombination $H_\star \sim  10^{-29}\,\mathrm{eV}$. These very small scales imply that the potential should be extremely flat, which in turn means that scalar fields relevant for birefringence measurements need to be very light. Indeed, explicit calculations show that only for $m_\phi \lesssim 10^{-25}\,\text{eV}$ are there open regions of parameter space in $g_{\phi \gamma \gamma}$ that can explain the measurement~\cite{Fujita:2020aqt}. In the particular case for which $\phi$ plays the role of dark energy, this corresponds to~\cite{Fujita:2020ecn}: $10^{-41}\,\text{eV} \lesssim  m_\phi \lesssim 10^{-34}\,\text{eV}$ with $10^{-12}\,\text{GeV}^{-1}\gtrsim g_{\phi\gamma\gamma} \gtrsim M_{\rm Pl}^{-1}$, where $M_{\rm Pl} = 1.22\times 10^{19}\,\text{GeV}$ is the Planck mass.

\secskip

\newpage

\emph{Global Symmetries and Light Scalar Fields.---} Global symmetries, even if only approximate, provide a way to understand the lightness of pseudoscalar fields. The reason for this is ultimately that, in the limit in which the symmetry is exact but spontaneously broken, the scalar field can be identified as a massless Goldstone boson~\cite{Goldstone:1961eq,Goldstone:1962es}. Beyond this, the explicit breaking of the symmetry will lead to a small but non-zero mass for $\phi$, which could arise from a number of sources. For example, the $U(1)$ symmetry invoked in the Peccei-Quinn mechanism is known to be explicitly broken by QCD gauge instantons~\cite{tHooft:1976rip}, but is also expected to be sensitive to additional gravitational sources of symmetry breaking~\cite{Kamionkowski:1992mf,Holman:1992us,Barr:1992qq,Ghigna:1992iv}. In the case of cosmic birefringence, regardless of the dominant source of global symmetry breaking, we expect that there will generically be some gravitational contribution. Any gravitational source should then be small enough such that the mass of $\phi$ does not exceed $m_\phi \sim 10^{-25}\,\mathrm{eV}$, allowing us to set an observational bound on the breaking of global symmetries due to gravity. 

In connection with typical axion-like potentials, we parametrise the breaking of global symmetries induced by gravity as~\cite{Marsh:2015xka}:
\begin{equation}\label{eq:DeltaV}
    \Delta V(\phi)_{\rm QG} = M_{\rm Pl}^4 e^{-S_{\rm QG}} [1- \cos\left(\phi/f\right)]\,,
\end{equation}
where we have normalised the potential to $M_{\rm Pl}$ as we are interested in gravitational effects. Here, $f$ is the energy scale at which the global symmetry would be spontaneously broken, and $S_{\rm QG}$ is a dimensionless parameter that quantifies the magnitude of the gravitational symmetry breaking -- often associated with the action of a non-perturbative object~\cite{Kallosh:1995hi}.

Given this parametrisation, we can expand the potential in Eq.~\eqref{eq:DeltaV} and estimate the contribution to the mass of $\phi$ from  gravitational effects to be:
\begin{equation}\label{eq:Deltam}
(\Delta m_\phi)_{\rm QG} \sim (M_{\rm Pl}^2/f) \, e^{-S_{\rm QG}/2}\,.
\end{equation}
This should not exceed $m_\phi \sim 10^{-25}\,\text{eV}$ in order for the dynamics of $\phi$ to be compatible with the cosmic birefringence measurement. By enforcing $(\Delta m_\phi)_{\rm QG} < 10^{-25}\,\text{eV}$, we can then set an observational upper bound on the magnitude of global symmetry breaking by gravity. This bound, written in terms of $S_{\rm QG}$, reads:
\begin{equation}\label{eq:SRequirement}
S_{\rm QG} > 2 \log\left(\frac{M_{\rm Pl}}{10^{-25}\,\text{eV}}  \right)+ 2 \log\left(\frac{M_{\rm Pl}}{f}\right) > 250\,,
\end{equation}
where in the last inequality we have used $f < M_{\rm Pl}$ from the general expectation that any new force should be stronger than gravity~\cite{ArkaniHamed:2006dz}. It can also be seen from Eq.~\eqref{eq:SRequirement} that the bound on $S_{\rm QG}$ is only logarithmically dependent on the mass of the scalar field $m_\phi$.

One should note that the bound on $S_\mathrm{QG}$ derived in Eq.~\eqref{eq:SRequirement} is dependent on the form of the gravitational corrections to the potential $\Delta V(\phi)_{\rm QG}$. In particular, in Eq.~\eqref{eq:DeltaV}, we have assumed that the explicit global symmetry breaking contributions are not protected by any gauge symmetry. This may not be fully generic, however the effect of this gauge protection can be characterised by an additional suppression factor, $(f/M_{\rm Pl})^n$, in $\Delta V(\phi)_\mathrm{QG}$. The exact power $n$ will of course depend upon the details of the model under consideration, and a corresponding bound could be derived in each case. On the other hand, in the context of our argument, we expect $f$ to lie somewhere between the GUT scale and the Planck scale and therefore $f/M_{\rm Pl} \gtrsim 10^{-3}$. Consequently, we do not anticipate that the possible protection by gauge symmetries will significantly alter our results for $S_\mathrm{QG}$. Similarly, in Eq.~\eqref{eq:DeltaV}, we have also assumed that there is no additional suppression of the global symmetry breaking coming from other effects, such as supersymmetry. In such a scenario, the prefactor $M_{\rm Pl}^4$ should be replaced by $M_{\rm Pl}^2 \Lambda_{\rm SUSY}^2$~\cite{Svrcek:2006yi,Arvanitaki:2009fg}. Since experimentally we know that $\Lambda_{\rm SUSY} \gtrsim 10\,\text{TeV}$, the resulting bound on $S_{\rm QG}$ in scenarios with supersymmetry suppression will only be relaxed to $S_{\rm QG}> 210$.

\secskip
\emph{Conclusions and Implications.---} We have derived a bound on the magnitude of global symmetry breaking by gravity, motivated by a recent detection of cosmic birefringence in Planck legacy data. The observational status of this detection is currently at the $2.4\sigma$ level, however, a similar analysis can be performed with other existing CMB data sets~\cite{Choi:2020ccd,Ade:2014afa,Ade:2014gua,Benson:2014qhw}, which may significantly strengthen this claim. In this letter, we have shown how such a measurement can be used to provide insight into the breaking of global symmetries by gravity. To do this, we have presumed that this measurement is explained by a very light pseudoscalar field coupled to electromagnetism. This is supported by the original description of the phenomenon in terms of light scalar fields in cosmology, as well as the prevalence of pseudo-Goldstone bosons in many theories beyond the Standard Model. 

If confirmed, this bound would have clear implications not only for the construction of quantum gravity theories, but also for the viability of many particle physics models. In terms of quantum gravity, a bound of this type would shape the vacuum structure of any proposed theory, including the action of any non-perturbative instantons breaking global symmetries -- such as microscopic black holes, wormholes or string instantons. To put this in context, the action of an axionic wormhole is $S_{\mathrm{QG}} \simeq M_{\rm Pl}/f$~\cite{Giddings:1987cg}. Taken at face value, our bound would therefore disfavour quantum gravity scenarios with wormholes that have $f > M_{\rm Pl}/250$.  Similarly, in string theory there are a variety of instantons breaking global symmetries~\cite{Svrcek:2006yi}, which typically have actions $S_\mathrm{QG} \lesssim 300$~\cite{Svrcek:2006hf}. A bound of the type derived above would then disfavour string theory realisations where all instantons have $S_{\rm QG} < 210$. 

String theories contain a large variety of axion-like particles, so it is interesting to consider the case that one of them is the QCD axion that solves the Strong CP problem. A minimal requirement for this to be viable is that at least one string instanton should have $S_{\rm QG} \gtrsim 190$~\cite{Svrcek:2006yi,Alvey:2020nyh}. By comparing this theoretical bound with the one derived from cosmic birefringence we can see that if our limit holds, then the Peccei-Quinn solution to the Strong CP problem will not be spoiled by gravitational effects. This is simply because additional symmetry breaking contributions will be many orders of magnitude smaller than those induced by QCD instantons. In addition, this would imply that if a string theory satisfies our bound, then it should automatically be able to contain a viable QCD axion, at least in principle. Of course, there are caveats to this comparison in that \emph{(i)} a given string theory does not necessarily have to contain a QCD axion and \emph{(ii)} while our bound is observational - being based on an experimental measurement - it has only a limited statistical significance so far, as discussed above. Nonetheless, this discussion highlights the potential of this sort of observational limit to connect particle physics phenomenology and quantum gravity theories.

To conclude then, confirming the existence of an almost exact global symmetry in nature would open an avenue to uncovering an important feature of quantum gravity. In this letter, we have suggested that the phenomena of cosmic birefringence could be the way to establish such an observational pathway.

\vspace{20pt}
\emph{Acknowledgements.---} M.E.A. is grateful to Gaurav Tomar for suggesting M.E.A. to present~\cite{Minami:2020odp} in a Journal Club. We thank Gonzalo Alonso \'Alvarez, Alejandro Ibarra, Mario Reig, Nash Sabti, and Sam Witte for comments on a draft version of this manuscript. J.A. is a recipient of an STFC quota studentship. M.E.A. is supported by a Fellowship of the Alexander von Humboldt Foundation.

\bibliography{biblio}
\end{document}